\documentclass[12pt,preprint]{aastex}

\slugcomment{accepted for publication in the ApJ Letter}

\shorttitle{Rotation-Resolved Spectroscopy of a Young Asteroid}
\shortauthors{N. Takato}

\begin{document}


\title{Rotation-Resolved Spectroscopy of a Very Young Asteroid, (1270) Datura}


\author{Naruhisa Takato\altaffilmark{1}}
\affil{Subaru Telescope, 650 North A`ohoku Pl., Hilo, HI 96720}
\email{takato@naoj.org}


\altaffiltext{1}{another affiliation: SOKEN-DAI, Hayama, Kanagawa 240-0193, Japan}


\begin{abstract}
(1270) Datura is the largest member of a very young asteroid cluster that was
 thought to be broken-up 0.45 Myr ago.  
 The light-curve and the rotation-resolved reflectance spectra 
 (0.6 $\mu$m -- 1.0 $\mu$m) were observed in order to find ``fresh'' surface. 
 Our data show no significant spectral variation along the rotation phase. The depth of the 0.95 $\mu$m absorption band, which indicates the degree of space weathering, was similar to that of an old S-type asteroid. This suggests that the reflectance spectrum in this wavelength range changes rapidly and saturates the depth of the 0.95 $\mu$m absorption in less than 0.45 Myr in the main belt environment.
\end{abstract}





\section{Introduction}

A parent body of an ordinary chondrite, one of the most common stony meteorites, is believed to be a S-type asteroid that is commonly distributed in the inner main belt region and is thought to consist of silicate rocks covered with regolith. However, there are discrepancies between the reflectance spectra of ordinary chondrites and S-type asteroids: S-type asteroids show redder spectral slope in the visible wavelength and the depth of the absorption band at $\sim0.95$ $\mu$m is shallower than that of ordinary chondrites \citep{cha73}. 

These spectral discrepancies are attributed to ``space weathering,'' which alter a fresh, chondritic spectra to the S-type asteroid spectra (e.g., Clark et al. 2002). Although details of the physical process of the space weathering have not been fully understood, it is thought that irradiation of high energy particles (solar wind or cosmic ray) and/or bombardment of micro-meteorites makes nanophase iron in the surface of regolith that cases the spectral change \citep{ada71, hap00, pie00, sas01}.

One of the unsolved issues on space weathering is to determine the time scale of the weathering process. Dynamical investigations of asteroid families are enabling us to create a time scale of the history of asteroid evolution (e.g., Nesvorn\'y et al. 2005). \citet{nes05} related the family age and the spectral slope, and \citet{wil08} revised their results and found that the time scale of the spectral slope change by space weathering was $570\pm220$ Myr.
However, \citet{cha07} and \citet{ver07} observed (832) Karin, whose age was estimated to be 5.8 Myr, and found that the depth of 0.95 $\mu$m absorption band was as shallow as a typical S-type asteroid. These observations suggest that the depth of 0.95 $\mu$m absorption band changes faster than the reddening of the spectral slope. 

Recently \citet{nes06a} found a very young asteroid cluster, Datura cluster, whose age was estimated to be $0.45\pm0.05$ Myr. Datura cluster has been exposed for only $1/10$ of the age of the Karin cluster. Thus this cluster is a good target to examine the alteration speed by space weathering at the 0.95 $\mu$m absorption band.

(1270) Datura has a diameter of about 11 km and is orbiting in the inner main belt region ($a=2.2$ AU). 
I have observed (1270) Datura, which is the largest member of the Datura cluster, expecting to find a ``fresh'' region on its surface because the space-weathered old surface was removed by the cluster-forming break-up and a fragment might expose a fresh, interior surface on the whole area or at least in some area. 
In this letter I will report the rotation-resolved reflectance spectra of (1270) Datura and its implication for the space weathering.

\section{Observations and Results}

\begin{figure}
\plotone{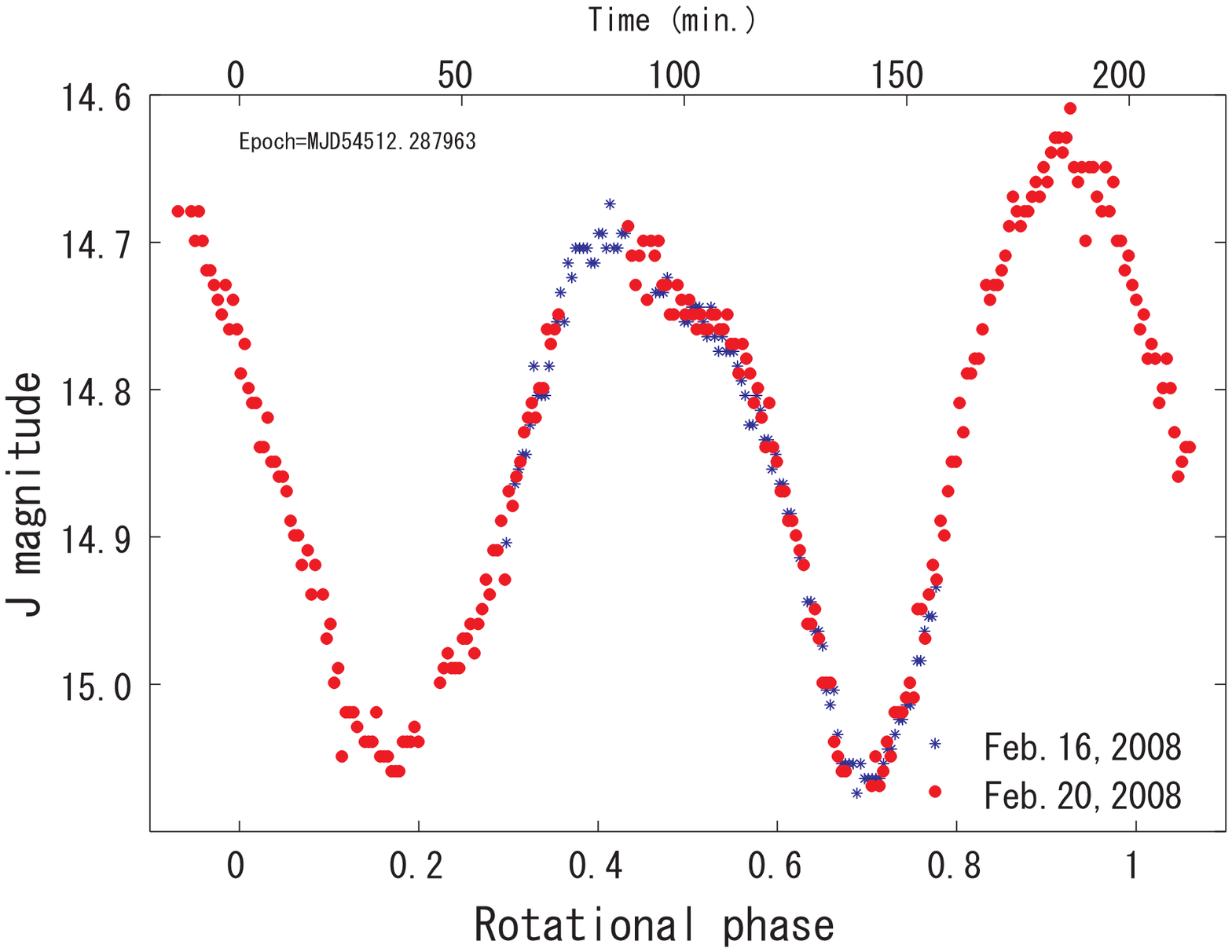}
\caption{$J-$band light-curve of (1270) Datura. Zero phase is chosen at MJD 54512.287963 (subtracted light traveling time). The sidereal period is estimated to be $P=3.359$ hr.
\label{fig_lc}}
\end{figure}

Light curve observations were made on 2008 February 16 and 20(UT) by using $J$-band imaging of MOIRCS \citep{ich06} on the Subaru telescope in order to determine the rotational phases at our spectral observations. Most of the exposure time were set to 30 sec.
Photometric standard star, FS132, was also observed.
The nights were photometric.
Figure~\ref{fig_lc} shows obtained $J$-band light curve of (1270) Datura. The sidereal rotation period was derived to be 3.359 hr and the amplitude was about 0.4 mag$_{pv}$, which are consistent with previous results \citep{sze05}.

Spectroscopic observations were made on 2008 February 26(UT) using the grism
mode of FOCAS \citep{kas02} on the Subaru telescope. The wavelength range was $\lambda=0.6$ $\mu$m -- 1.0 $\mu$m and the slit width was set to 1.0 arcsec, which resulted the spectral resolution to be R=400. The position angle of the slit was set to the direction of the asteroid motion.
The sky was clear and the spectra of (1270) Datura were taken continuously and covered about $3/4$ of a full rotational phase with 26 spectra. Exposure time of each spectrum was 180 sec. Spectra of a solar analogue star HD60298(G2V) were taken during the observation to compensate atmospheric absorption features and to derive reflectance spectra of (1270) Datura.

\begin{figure}
\epsscale{0.6}
\plotone{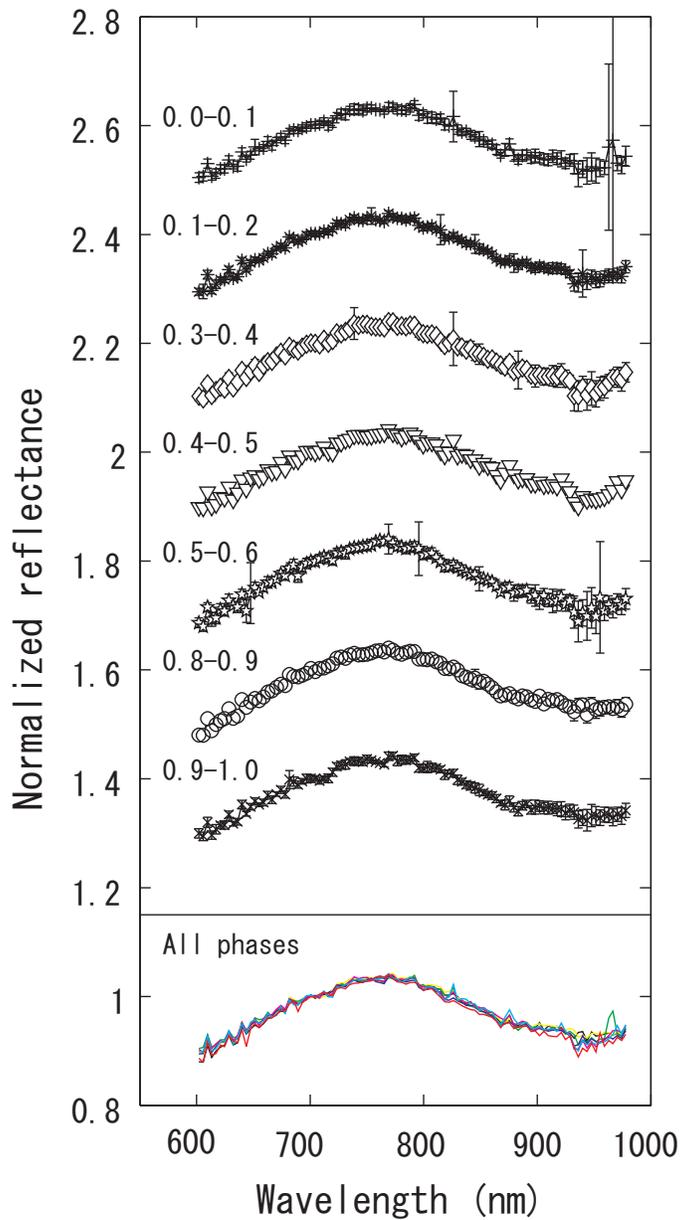}
\caption{Average reflectance spectra of (1270) Datura for seven different 0.1 bin of rotational phase. Each spectrum is normalized to unity at 700 nm and shifted vertically by 0.2 for clarity. Data near 760 nm are excluded due to strong telluric  O$_2$ absorption band. At the bottom panel, all the spectra are plotted with no vertical offset to see spectral difference among various rotational phases. \label{fig_sp}}
\end{figure}

The spectra were traced and extracted using IRAF package and then were divided by the spectra of the solar analogue star observed at a similar air mass to (1270) Datura.
Figure~\ref{fig_sp} shows the obtained reflectance spectra of (1270) Datura averaged at several rotational phases.
No significant spectral variations were detected along the rotational phases.





\section{Discussion}

The rotation curve shows fairly large amplitude that means we are not seeing (1270) Datura from above its pole, thus we see most of its surface when observing nearly a full rotational period.
The fact that there is no significant difference among the spectra obtained at different rotation phases indicates that the surface of (1270) Datura has no significant large scale spatial variation in composition and in the degree of space weathering. The rotation speed of (1270) Datura is slow enough, and the size is large enough to be covered with regolith that must be newly created at the cluster-forming collision at 0.45 Myr ago and then deposited onto the surface. Since the deposition time scale might be longer compared to the rotation period of (1270) Datura, the dusts/gravels covered evenly over the surface.
Thus it is natural that the surface of (1270) Datura is almost homogeneous when observed in a large spatial scale as discussed by \citet{nes06b} and \citet{ver07} for (832) Karin.

\begin{figure}
\plotone{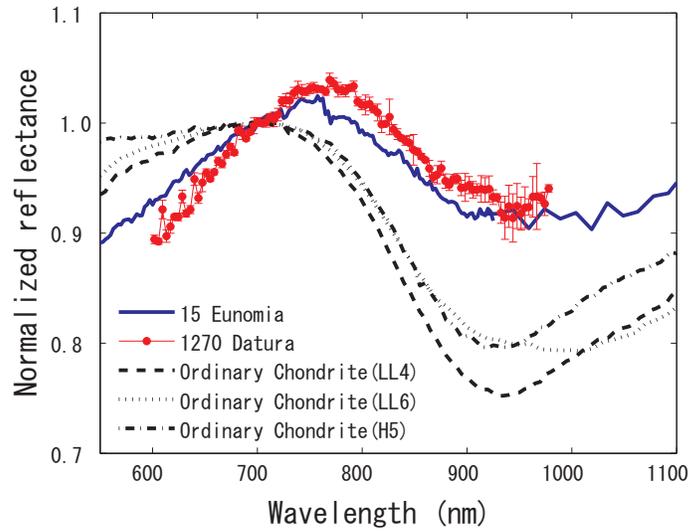}
\caption{Average spectra of (1270) Datura compared to old ($\sim2.5$ Gyr) S-type asteroid (15) Eunomia \citep{bus02,bur02} and ordinary chondrites (RELAB data base). The spectrum of (1270) Datura does not resemble that of ordinary chondrites, but is similar to an old S-type asteroid.\label{fig_comp}}
\end{figure}

To see the degree of space weathering of the surface of (1270) Datura, the phase-averaged spectrum is compared with that of an old S-type asteroid (15) Eunomia (estimated age is 2.5 Gyr \citep{nes05}) and some of the ordinary chondrites as shown in Figure~\ref{fig_comp}.
Contrary to our expectation, the spectrum of (1270) Datura does not resemble those of ordinary chondrites, but is close to an old S-type asteroid (15) Eunomia.
This means that the surface of (1270) Datura was already altered in the same degree with an old S-type asteroid in respect to the depth of 0.95 $\mu$m absorption band.

Because the surface was renewed 0.45 Myr ago by a cluster-forming collision, the change of the reflectance spectrum (from ordinary-chondrite-like one to that of S-type asteroids) must be completed in less than 0.45 Myr, if the age estimate by \citet{nes06a} 
and the concept of space weathering are
correct.


\citet{bin96, bin01} found that many of the near-earth objects (NEOs) are Sq- or Q-type, which connects S-type spectra to ordinary chondrite spectra. If NEOs have been space weathered in the same manner as (1270) Datura, the surface of those objects have been exposed much less than $0.45$ Myr. If the surface was exposed by a recent collision in the main belt, we can estimate the current collision rate in the main belt. 
However, we must be careful that the environment of space weathering in NEO region differs from the inner main belt. Also the establishment and stability of regolith might be different since the size of the objects of \citet{bin96, bin01} are one order-of-magnitude smaller than (1270) Datura.



\acknowledgments

I am grateful to T. Hattori for his support of observation and data 
reductions. I thank C. R. Chapman for his helpful comments and suggestions as a reviewer and S. Sasaki for his useful comments. 
Light-curve data were obtained during telescope engineering time of the Subaru telescope.



{\it Facilities:} \facility{Subaru (FOCAS, MOIRCS)}.

\clearpage



\begin{thebibliography}{}
\bibitem[Adams \& McCord (1971)]{ada71} Adams, J. B. \& McCord, T. B. 1971, Science, 171, 567
\bibitem[Binzel et al.(1996)]{bin96} Binzel, R. P., Bus, S. J., Burbine, T. H., Sunshine, J. M. 1996, Science, 273, 946
\bibitem[Binzel et al.(2001)]{bin01} Binzel, R. P., Harris, A. W., Bus, S. J., Burbine, T. H. 2001, Icarus, 151, 139
\bibitem[Burbine \& Binzel (2002)]{bur02} Burbine, T. H. \& Binzel, R. P. 2002, Icarus, 159, 468
\bibitem[Bus \& Binzel (2002)]{bus02} Bus, S. J. \& Binzel, R. P. 2002, Icarus, 158, 146
\bibitem[Chapman \& Salisbury (1973)]{cha73} Chapman, C. R. \& Salisbury, J. W. 1973, Icarus, 19, 507
\bibitem[Chapman et al.(2007)]{cha07} Chapman, C. R., Enke, B., Merline, W. J., Tamblyn, P., Nesvorn\'y, D., Young, E. F., Olkin, C., 2007, Icarus, 191, 323
\bibitem[Clark et al.(2002)]{cla02} Clark, B. F., Hapke, B., Pieters, C., Britt, D. 2002, Asteroids III, Bottke, W. F., Cellino, A., Paolicchi P., Binzel, R. P., Univ. of Arizona Press: Tucson, 585
\bibitem[Hapke (2000)]{hap00} Hapke, B. 2000, Lunar Planet. Sci., 31, 1087
\bibitem[Ichikawa et al.(2006)]{ich06} Ichikawa, T. et al. 2006, Proc. SPIE, 6269, 38
\bibitem[Kashikawa et al.(2002)]{kas02} Kashikawa, N., et al. 2002, PASJ, 54, 819
\bibitem[Nesvorn\'y et al.(2005)]{nes05} Nesvorn\'y, D., Jedicke, R., Whiteley, R. J., Ivezi\'c, $\check{\rm Z}$. 2005, Icarus, 173, 132
\bibitem[Nesvorn\'y et al.(2006a)]{nes06a} Nesvorn\'y, D., Vokrouhlick\'y, D., Bottke, W. F. 2006a, Science, 312, 1490
\bibitem[Nesvorn\'y et al.(2006b)]{nes06b} Nesvorn\'y, D., Enke, B., Bottke, W. F., Durda, D., Asphaug, E., Richardson, D. 2006b, Icarus, 183, 311
\bibitem[Pieters et al. (2000)]{pie00} Pieters, C. M., Taylor, L. A., Noble, S. k., Lindsay, P. K., Hapke, B, Morris, R. V., Allen, C. C., McKay, D. S., Wentworth, S. 2000, Meteorit. Planet. Sci., 35, 1101
\bibitem[RELAB(2008)]{rel08} RELAB data base, NASA RELAB facility at Brown Univ. (\url{http://www.planetary.brown.edu/relabdocs/relab.htm})
\bibitem[Sasaki et al.(2001)]{sas01} Sasaki, S., Nakamura, K., Hamabe, Y., Kurahashi, E., Hiroi, T. 2001, Nature, 410, 555
\bibitem[Sz\'ekely et al.(2005)]{sze05} Sz\'ekely, P. et al. 2005, Planet. Space Sci., 53, 925
\bibitem[Vernazza et al.(2007)]{ver07} Vernazza, P., Rossi, A., Birlan, M., Fulchignoni, M., Nedelcu, A., Dotto, E. 2007, Icarus, 191, 330
\bibitem[Willman et al.(2008)]{wil08} Willman, M., Jedicke, R., Nesvorn\'y, D., Moskovitz, N., Ivezi\'c, $\check{\rm Z}$., Fevig, R. 2008, Icarus, 195, 663
\end{thebibliography}
\end{document}